\documentclass[preprint,showpacs,preprintnumbers,amsmath,amssymb,nofootinbib]{revtex4}

\usepackage{graphicx}
\usepackage{bm}

\newcommand{\be}{\begin{eqnarray}}
\newcommand{\ee}{\end{eqnarray}}

\catcode`\@=11
\def\@versim#1#2{\vcenter{\offinterlineskip
\ialign{$\m@th#1\hfil##\hfil$\crcr#2\crcr\sim\crcr } }}
\catcode`\@=12
\begin{document}
\def\descriptionlabel#1{\bf #1\hfill}
\def\description{\list{}{%
\labelwidth=\leftmargin
\advance \labelwidth by -\labelsep
\let \makelabel=\descriptionlabel}}
\newcommand{\ds}{\displaystyle}
\newcommand{\lt}{\left}
\newcommand{\rt}{\right}
\newcommand{\no}{\nonumber}
\newcommand{\nn}{\nonumber \\}
\newcommand{\ov}[1]{\overline{#1}}
\newcommand{\eq}[1]{Eq.~(\ref{#1})}
\newcommand{\eqsand}[2]{Eqs.~(\ref{#1}) and (\ref{#2})}
\newcommand{\eqsto}[2]{Eqs.~(\ref{#1}--\ref{#2})}
\newcommand{\imag}{\mathrm{Im}\,}
\newcommand{\real}{\mathrm{Re}\,}
\newcommand{\gev}{\,\mbox{GeV}}
\newcommand{\mev}{\,\mbox{MeV}}
\newcommand{\sgn}{\mbox{sign}\,}
\newcommand{\dis}{\displaystyle}
\newcommand{\lsim}{\stackrel{<}{_\sim}}
\newcommand{\gsim}{\stackrel{>}{_\sim}}

\newcommand{\Bbar}{\,\overline{\!B}}
\newcommand{\gqtf}{{\Gamma (B(t) \rightarrow f )}}
\newcommand{\gqbtf}{{\Gamma (\Bbar{}(t) \rightarrow f )}}
\newcommand{\gqtfb}{{\Gamma (B(t) \rightarrow \ov{f} )}}
\newcommand{\gqbtfb}{{\Gamma (\Bbar{}(t) \rightarrow \ov{f} )}}
\newcommand{\bbd}{\ensuremath{B_d\!-\!\Bbar{}_d\,}}
\newcommand{\bbs}{\ensuremath{B_s\!-\!\Bbar{}_s\,}}
\newcommand{\bbq}{\ensuremath{B_q\!-\!\Bbar{}_q\,}}
\newcommand{\bbms}{\bbs\ mixing}
\newcommand{\bbmd}{\bbd\ mixing}
\newcommand{\BsorBsbar}{\raisebox{7.7pt}{$\scriptscriptstyle(\hspace*{8.5pt})$}
 \hspace*{-10.7pt}\!\Bbar_{s}}
\newcommand{\nuornubar}{
 \,\raisebox{5.5pt}{$\scriptscriptstyle(\hspace*{6.3pt})$}
 \hspace*{-7.8pt}\!\ov{\nu}}
\newcommand{\bra}[1]{\ensuremath{\langle #1 |}}
\newcommand{\ket}[1]{\ensuremath{| #1 \rangle }}
\newcommand{\fig}[1]{Fig.~\ref{#1}}
\newcommand{\lqcd}{\Lambda_{\textit{\scriptsize{QCD}}}}
\newcommand{\lbar}{\ov{\Lambda}}
\newcommand{\dm}{\ensuremath{\Delta M}}
\newcommand{\dg}{\ensuremath{\Delta \Gamma}}
\newcommand{\epm}[2]{
\raisebox{-0.5ex}{\shortstack[l]{$\scriptstyle+#1$\\$\scriptstyle-#2$}}}

\def\thefootnote{\fnsymbol{footnote}}
\begin{flushright}
KANAZAWA-09-08  \\
July, 2009
\end{flushright}
\vspace*{2cm}
\begin{center}
{\LARGE\bf Testing the new CP phase 
in a Supersymmetric Model with $Q_6$ Family Symmetry
by  B$_s$ Mixing}\\
\vspace{1.5 cm}

{\Large Kenji Kawashima}
\footnote[1]{e-mail:~kenji@hep.s.kanazawa-u.ac.jp}, {\Large Jisuke Kubo}
\footnote[2]{e-mail:~jik@hep.s.kanazawa-u.ac.jp}
\vspace {0.5cm}\\
{\it Institute for Theoretical Physics, Kanazawa University,\\
        Kanazawa 920-1192, Japan}
\vspace {0.5cm}\\
and 
\vspace {0.5cm}\\
{\Large Alexander Lenz}
\footnote[3]{e-mail:~alexander.lenz@physik.uni-regensburg.de}
\vspace {0.5cm}\\
{\it Institut f{\"u}r Theoretische Physik, Universit{\"a}t Regensburg,\\
        D-93040 Regensburg, Germany}\\
\end{center}
\vspace{1cm}
{\Large\bf Abstract}\\
The new contribution to
the non-diagonal matrix element $M_{12}$ 
of the neutral $B_s$ meson system  is investigated in a supersymmetric extension of the standard model
 based on the discrete $Q_6$ family symmetry.
We assume that CP is explicitly, but softly broken only
by the b terms in the soft supersymmetry breaking sector.
We find that the new contributions to $M_{12}$ are real, and 
that nevertheless there exists an observable  difference in  the CP 
phase compared with the standard model.
We focus our attention on  the flavor-specific CP asymmetry $a_{fs}^s$,
and find that $a_{fs}^s$ of the model
 is mostly negative and its size can be one order of magnitude larger the
 standard model value.
This prediction is consistent with the current experimental value, and 
can be experimentally tested in the  near future.

\newpage
\setcounter{footnote}{0}
\def\thefootnote{\arabic{footnote}}

\section{Introduction}
Recent experimental data from TeVatron, see e.g.~\cite{Paulini:2009ri}
give some hints for possible deviations from the standard model (SM)  
in the $B_s$ mixing system.
In the standard model the mixing of the neutral mesons is described 
by the famous box-diagrams. The dispersive part of these diagrams 
is denoted by $M_{12}$, it is due to heavy internal particles and 
therefore sensitive to possible new physics contributions. The 
absorptive part of the box-diagrams - denoted by $\Gamma_{12}$ - is 
due to light internal particles and it can not be affected by large 
new physics contributions, see e.g. \cite{LN2006} for more details.

The phases of $M_{12}$ and $\Gamma_{12}$ alone are unphysical, but 
the phase difference can be measured. We use the definition
$\phi_s = \mbox{arg} (-M_{12} / \Gamma_{12} )$. $|M_{12}|$, $|\Gamma_{12}|$
and $\phi_s$ can be related to the following observables:
the mass difference $\Delta M_s = 2 |M_{12}|$ was measured at
CDF \cite{DMCDF} and D0 \cite{DMD0}. HFAG \cite{HFAG} combines the numbers to
$\Delta M_s = 17.78 \pm 0.12 \, \mbox{ps}^{-1}$.
From the angular analysis in the decay $B_s \to J/\psi \phi$ one can extract
the decay rate difference $\Delta \Gamma_s = 2 |\Gamma_{12}| \cos (\phi_s)$
and the mixing phase 
$\beta_s = -\mbox{arg} (- V_{ts}^* V_{tb}/V_{cs}^* V_{cb})$, 
c.f. \cite{FDN}. 
The standard model predicts very small numerical values for the mixing phases,
$\beta_s \approx (2.2 \pm 0.6)^\circ$ and $\phi_s \approx
 (0.24 \pm 0.04)^\circ $ \cite{LN2006}.
If new physics contributes to $M_{12}$, then $\phi_s$ and $-2 \beta_s$ 
are shifted by the same value, which we denote by $\phi_s^\Delta$,
 see the note added in \cite{ELBA} for
more details. Currently both CDF \cite{Aaltonen:2007he}
and D0 \cite{2008fj} did tagged analyses of the decay $B_s \to J/\psi \phi$ 
and they obtain values for the mixing phase  which differ about 2.2 
$\sigma$ \cite{HFAG} from the SM.
Similar deviations are obtained by CKMfitter \cite{CKMfitter}, 
while UTfit \cite{UTfit} sees a slightly bigger
effect.

Finally we can relate the box diagrams to flavor-specific CP asymmetries, 
which are also called 
semileptonic CP-asymmetries: $ a_{fs} = $ Im $(\Gamma_{12}/M_{12})  
= (\Delta \Gamma / \Delta M)  \tan \phi_s$.
Theses asymmetries can be extracted directly from experiment \cite{asldirect} 
or they can be derived from the
di-muon asymmetry \cite{asldimuon}. The standard model expectation for the semileptonic
CP asymmetry in the $B_s$ system is again very small, 
$a_{fs}^s \approx (2.06 \pm 0.57) \cdot 10^{-5}$ 
\cite{LN2006}. Currently the experimental uncertainties in $a_{fs}^s$ 
are still much larger than the standard 
model value.
If the particular strong suppression pattern of the standard model for 
$\phi_s$ and $a_{fs}^s$ 
is not present in a new physics extension, then these quantities might be 
enhanced considerably (up
to a factor of 250, see \cite{LN2006}). In order to distinguish new physics 
effects from hadronic uncertainites,
precise standard model predictions are mandatory. We take the numerical 
values for the standard model expectations
from \cite{LN2006}, which uses results of 
\cite{BBD06,BBGLN98, BBGLN02, BBLN03,rome03}.

In this letter we  consider a supersymmetric extension
of the SM based on the discrete $Q_6$ family symmetry
\cite{Babu:2004tn,Kajiyama:2005rk},
and  investigate the extra contribution to
$M_{12}$. 
In \cite{Kifune:2007fj} we have stressed a minimal content of the Higgs
multiplets, i.e. no extra Higgs multiplet that is $SU(2)_L\times U(1)_Y$
singlet. 
We have then found that
it is possible, without contradicting renormalizability, to have
the one + two structure for each family. By the one + two structure for a family
we mean
a family (including the Higgs sector) with three family members;
one member in the $Q_6$ singlet representation 
and the other two in the $Q_6$ doublet representation.
As in \cite{Kifune:2007fj} we assume that CP is explicitly, but softly broken only
by the b terms in the soft supersymmetry breaking sector.
Therefore, all  other parameters of the model are real.
We take into account the contribution  to $M_{12}$ coming from the supersymmetry 
breaking sector
as well as  from the exchange of the flavor-changing neutral Higgs bosons.
It turns out that both contributions are real, and 
that nevertheless there exists an observable  difference in the CP 
phase in the mixing of the neutral mesons.
Specifically, we focus our attention on  
the extra phase $\phi_s^\Delta$ and the flavor-specific CP asymmetry $a_{fs}^s$,
because they are
accidentally very small in the SM.
We find that $a_{fs}^s$ of the model is mostly
negative and can be  one order of magnitude larger the SM value
in size.

\section{The  model}
The model is briefly described below (the details of the model
can be found in \cite{Kajiyama:2005rk,Kifune:2007fj}).
The $SU(2)_L$ doublets
of the quark and Higgs supermultiplets
are denoted by $Q $ and  $ H^u, H^d$, 
respectively. 
Similarly, $SU(2)_L$ singlets 
of the quark supermultiplets are denoted by
$U^c$ and $D^c$. (Here we restrict ourselves to
 the quark sector. The prediction in the lepton sector,
 which is given in \cite{Kajiyama:2005rk},
is  the same as in the $S_3$ model of \cite{Kubo:2003iw,Mondragon:2008gm}.)
The $Q_6$ assignment is shown in Table \ref{assignment},
where we assume $R$ parity.
\begin{table}
\begin{center}
\begin{tabular}{|c|c|c|c|c|c|c|c|c|c|c|c|}
\hline
 & $Q$ 
 & $Q_3$  
& $U^c,D^c$  
& $U^c_3,D^c_3$ 
 & $L$ & $L_3$ 
 &$E^c,N^c$ & $E_3^c$  &   $N_3^c$ 
  & $H^u,H^d$
 & $H^u_3,H^d_3$ 
\\ \hline
$Q_6$ &${\bf 2}_1$ & ${\bf 1}_{+,2}$ &
 ${\bf 2}_{2}$ &${\bf 1}_{-,1}$ &${\bf 2}_{2}$ &
${\bf 1}_{+,0}$  & ${\bf 2}_{2}$ & 
${\bf 1}_{+,0}$ & ${\bf 1}_{-,3}$ &
${\bf 2}_{2}$ & ${\bf 1}_{-,1}$  \\ \hline
\end{tabular}
\caption{ \footnotesize{The $Q_{6}$ assignment 
of the chiral matter supermultiplets.
The group theory notation is given in 
 Ref.~\cite{Babu:2004tn}. For completeness we show the
 $Q_{6}$ assignment of the leptons, too.}}
\label{assignment}
\end{center}
\end{table}
In what follows
we discuss successively  the Yukawa sector, the supersymmetry
breaking sector and the Higgs sector.
The crucial observation of \cite{Kifune:2007fj} 
in achieving the minimality of the Higgs sector is that 
softly-broken supersymmetry allows for
each sector of the model to have certain own symmetries
without loosing  renormalizability.  
Table \ref{symmetry} shows the symmetry structure 
used in \cite{Kifune:2007fj},
where the symbols are explained in the caption.
\vspace{0.5cm}
\begin{table}
\begin{center}
\begin{tabular}{|c|c|c|c|c|}
\hline
 & ${\bf Y},{\bf h}$ & 
 ${\bf m}$
 & $\mu $  sector & $b$ terms
 \\ \hline
$Q_6$ & $\bigcirc$ & $\bigcirc$ & $\times$  &  $\times$
\\ \hline
$O_2$ & $\times$ &$\bigcirc$ & $\bigcirc$  &  $\times$
\\ \hline
$Z_2$ & $\times$ & $\bigcirc$ & $\bigcirc$  &  $\bigcirc$
\\ \hline
CP & $\bigcirc$ & $\bigcirc$ & $\bigcirc$  &  $\times$
\\ \hline
R & $\bigcirc$ & $\bigcirc$ & $\bigcirc$  &  $\bigcirc$
\\ \hline
\end{tabular}
\caption{ \footnotesize{The symmetry of the different sectors.
${\bf Y}, {\bf h}$ and 
 ${\bf m}$ stand for 
 the Yukawa,  tri-linear
 and   soft scalar mass sector, respectively.
 $O_2$ in the soft scalar mass sector is accidental.
$Z_2$ is a subgroup of $O_2$.
CP is explicitly, but softly  broken only by the $b$ terms.
All the symmetries are compatible with each other,
and consequently, the model is renormalizable.}
 }
\label{symmetry}
\end{center}
\end{table}

\subsection{The Yukawa sector and the CKM parameters}
The superpotential for  the Yukawa interactions in
the quark sector is given by
\begin{eqnarray}
W_q
&=&Y_a^u Q_3 H_3^u U_3^c + Y_b^u (Q_1 H_2^u+Q_2 H_1^u) U_3^c
+ Y_{b'}^ u Q_3 (H_1^u U_2^c-H_2^u U_1^c)\nonumber\\
&&+ Y_c^u (Q_1 U_2^c+Q_2 U_1^c)H_3^u\nonumber\\
&&+Y_a^d Q_3 H_3^d D_3^c + Y_b^d (Q_1 H_2^d+Q_2 H_1^d) D_3^c
+ Y_{b'}^ d Q_3 (H_1^d D_2^c-H_2^d D_1^c)\nonumber\\
&&+ Y_c^d (Q_1 D_2^c+Q_2 D_1^c)H_3^d\ .
\label{yukawa1}
\end{eqnarray}
All the Yukawa couplings are  real.
So, the VEVs of the Higgs fields have to be complex
to obtain the CP phase of the CKM matrix.
Thanks to the $Z_2$  invariance
of  the scalar potential (see (\ref{scalarp1}))
under
\be
H^{u,d}_+&=&\frac{1}{\sqrt{2}}
(H^{u,d}_1+H^{u,d}_2)
\rightarrow H^{u,d}_+~,~
H^{u,d}_- =\frac{1}{\sqrt{2}}
(H^{u,d}_1-H^{u,d}_2) \rightarrow -H^{u,d}_-,
\label{Hpm}
\ee
the VEVs
\footnote{Fields with a hat are the scalar components
of the corresponding superfields.}
\be
<\hat{H}_{-}^{0u,d}> &=& 0, <\hat{H}_{+}^{0u,d}>
 =\frac{v_{+}^{u,d}}{\sqrt{2}}\exp i \theta_+^{u,d}, 
<\hat{H}_{3}^{0u,d}> 
=\frac{v_3^{u,d}}{\sqrt{2}}\exp i \theta_3^{u,d}
\label{vev1}
\ee
can become a local minimum, where we assume that 
$v_{+}^{u,d}$ and $v_{3}^{u,d}$  are real and positive.
(The Yukawa interactions  do not respect the $Z_2$ symmetry,
but due 
to the $Q_6$ family symmetry they can not induce  $Z_2$-violating scalar potential terms
of dimension less than or equal to four
in higher orders in perturbation theory.)
From the Yukawa interactions (\ref{yukawa1})
along with the  form of the VEVs (\ref{vev1})
we obtain the fermion mass matrices.
 In diagonalizing these mass matrices 
  we found \cite{Babu:2004tn} that the CKM mixing matrix
 can be written
 as 
 \be
   V_{\rm CKM} &=& (U^{u}_L)^\dag  U_L^d=
 O^{uT}_L P_u^\dag P_d O_L^d,
\label{ckm}
 \ee
 where  $O^{u}_L$ and  $O_L^d$ are orthogonal matrices, 
 and
 \be
 P_{u,d}  &=&
\frac{1}{\sqrt{2}}\left( \begin{array}{ccc}
1 & \exp (i2\Delta \theta^{u,d}) & 0\\-1 & 
\exp (i2\Delta \theta^{u,d})& 0\\
 0 & 0 &\sqrt{2} \exp (i\Delta \theta^{u,d})
\end{array}\right)~,~\Delta \theta^{u,d} =
 \theta_3^{u,d}-\theta^{u,d}_+,\\
P_q  &=& P_u^\dag P_d=\mbox{diag.}~(
1, \exp (i2\theta_q), \exp (i\theta_q))~,~
\theta_q =\theta^u_+-\theta^d_+-\theta^u_3
+\theta^d_3.
\label{phase}
\ee
There are nine independent theory parameters, i.e.,
$Y_a^{u,d} v_3^{u,d},
Y_c^{u,d} v_3^{u,d},Y_b^{u,d} v_+^{u,d},
Y_{b'}^{u,d} v_+^{u,d}$ and $\theta_q$, 
to describe the CKM parameters. So, there is one prediction
which can be displayed in different planes.
They are presented in \cite{Araki:2008rn}.
 
Since the purpose of the present paper is
to calculate the observable CP phases in the
$B^0$ mixing, it is sufficient to
consider a single point in the space of the theory parameters.
So, throughout this paper we use the following
theoretical values  \cite{Kifune:2007fj}:
\be
m_u/m_t &=& 0.766\times 10^{-5},
m_c/m_t=4.23 \times 10^{-3},
m_d/m_b=0.895 \times 10^{-3},
m_s/m_b=1.60 \times 10^{-2},
\nn
|V_{\rm CKM}| &=&
\left( \begin{array}{ccc}
0.9740 & 0.2266  & 0.00362
\\  0.2265   & 0.9731& 0.0417
  \\ 0.00849 &0.0410& 0.9991
\end{array}\right),~
|V_{td}/V_{ts}| =0.207~,
\label{predQ}\\
\sin 2\beta (\phi_1) &=&0.690~,~\gamma(\phi_3)=63.4^o~.
\label{ckm-parameters}
\ee

\subsection{Soft-supersymmetry-breaking sector
and the phase alignment}
As we can see from Table \ref{symmetry} the tri-linear couplings ${\bf h}$ and
soft scalar mass terms ${\bf m}$ have the same family symmetry 
as  the Yukawa sector.
Consequently, the tri-linear couplings and the
soft scalar mass matrices have the following form:
\be
{\bf \tilde{m}^2}_{aLL}& =&
{m}^2_{\tilde{a}L}~ \mbox{diag.}~
(a_{L}^{a}~,~ a_{L}^{a}~,~ b_{L}^{a})~~(a=q,l)~,\\
{\bf \tilde{m}^2}_{aRR}& =&
{m}^2_{\tilde{a}R}~ \mbox{diag.}~
(a_{R}^{a}~,~ a_{R}^{a}~,~ b_{R}^{a})~~(a=u,d,e)~,
\label{scalarmass}\\
\left({\bf \tilde{m}^2}_{aLR}\right)_{ij} 
&=&
A_{ij}^a\left( {\bf m}^a \right)_{ij}=
\tilde{A}_{ij}^a ~\sqrt{{m}_{\tilde{a}L}{m}_{\tilde{a}R}}~\left( {\bf m}^a \right)_{ij} 
~~(a=u,d,e)~,
\label{Aterm}
\ee
where ${m}_{\tilde{a}L,R}$ denote the average of the  squark 
and slepton masses, respectively,   $(a_{L(R)}^a, b_{L(R)}^a)$ are
dimensionless free real parameters of $O(1)$, 
 $A_{ij}^{a}$ are free parameters of dimension one,
 and ${\bf m}^a$ are the fermion mass matrices.
Note that $a_{L,R}^a$ and $A_{ij}^a$ are all real, 
because we impose CP invariance in the tri-linear sector as well
as in the soft-scalar mass sector.

The quantities \cite{Hall:1985dx,Gabbiani:1988rb}
\be
\delta_{LL(RR)}^{a} &=&
U_{aL(R)}^{\dagger} ~{\bf \tilde{m}^2}_{aLL(RR)}~
 U_{aL(R)}/{m}^2_{\tilde{a}}~\mbox{and}~
\delta_{LR}^{a} =
U_{aL}^{\dagger}~ 
{\bf \tilde{m}^2}_{aLR} ~U_{aR}/{m}^2_{\tilde{a}}
\label{Delta1}
\ee
in the super CKM basis  are used widely to parameterize FCNCs and CP violations coming
from the soft supersymmetry breaking   sector, where the unitary matrices 
$ U$'s to rotate the fermions to the mass
eigenstates  are given in \cite{Kifune:2007fj}.
The imaginary parts of
$\delta$'s
contribute to CP violating processes induced  in the soft supersymmetry breaking  sector.
Recall that  the phases of
${\bf m}^2_{aLR}$ can come only from the complex VEVs (\ref{vev1}).
As we can see from  (\ref{phase}) the unitary matrices have the form
$U_L^{u,d}= P_{u,d} O_L^{u,d}$, where only $P_{u,d}$  are
complex. Since $P_{u,d}$  commute with
${\bf m}^2_{aLL,RR}$ (because their first $2\times 2$ block is
proportional to the identity matrix), $\delta_{LL,RR}^a$ have no
imaginary part.  Further, ${\bf m^2}_{aLR}$ has
the same phase structure as the corresponding fermion mass matrix
$ {\bf m}^a$, and it turns out that $\delta_{LR}^a$, too, are real. So,
the imaginary part of
$(\delta_{12,21,13,31,23,32}^d)_{LL,RR,LR,RL}$ which would contribute
to $\mbox{Im} M_{12}^{\rm new}$ is absent.
Therefore, as far as the soft scalar masses and the
left-right soft masses in the soft-supersymmetry-breaking sector
are concerned, there is no extra CP violating phase.

 In Table \ref{deltas} we show the actual values of the $\delta$'s
 which should be compared with the experimental bounds
 \footnote{See \cite{Kajiyama:2005rk} and references therein.}.
 These constraints come from the mass differences
 of the neutral mesons, i.e.,
 $\Delta M_K, \Delta M_d$ and $ \Delta M_s$.
We see that no fine tuning of the soft-supersymmetry
breaking parameters is needed to satisfy the experimental
constraints.
These  contributions from the supersymmetry breaking sector
should be added to the contribution coming from
the exchange of the flavor-changing neutral Higgs bosons.
In the best situation one can have they cancel each other.
As we will see, even in this situation, that is, even if we assume that
the   contributions from the supersymmetry breaking sector
can be freely chosen, we are able to make predictions
on the CP violating quantities such as the flavor-specific CP
asymmetry. 

\begin{table}[thb]
\begin{center}
\begin{tabular}{|c||c|c|}
 \hline
 &  Exp. bound  & $Q_6$ Model \\
   \hline  \hline
$\sqrt{|\mbox{Re}(\delta^d_{12})^2_{LL,RR}|}$
& $4.0 \times 10^{-2} ~\tilde{m}_{\tilde{q}}$
& $(LL)1.2 \times 10^{-4} \Delta a_L^q,(RR)1.7 \times 10^{-1}\Delta a_R^d$
\\ \hline
 $\sqrt{|\mbox{Re}(\delta^d_{12})_{LL}(\delta^d_{12})_{RR}|}$
& $2.8 \times 10^{-3} ~\tilde{m}_{\tilde{q}}$
&$4.5 \times 10^{-3}\sqrt{\Delta a_L^q \Delta a_R^d}$
\\ \hline
$\sqrt{|\mbox{Re}(\delta^d_{12})^2_{LR}|}$
 & $4.4 \times 10^{-3} ~\tilde{m}_{\tilde{q}}$
&$\sim 2 \times 10^{-5}(\tilde A^d_a-\tilde A^d_b-\tilde A^d_{b'}
+\tilde A^d_c)\tilde{m}_{\tilde{q}}^{-1}$
\\ \hline
 $\sqrt{|\mbox{Re}(\delta^d_{13})^2_{LL,RR}|}$
 & $9.8 \times 10^{-2} ~\tilde{m}_{\tilde{q}}$
&$(LL)7.8 \times 10^{-3}\Delta a_L^q,(RR)1.4 \times 10^{-1}\Delta a_R^d$
\\ \hline
$\sqrt{|\mbox{Re}(\delta^d_{13})_{LL}(\delta^d_{13})_{RR}|}$
&  $1.8 \times 10^{-2} ~\tilde{m}_{\tilde{q}}$
&$3.4 \times 10^{-2}\sqrt{\Delta a_L^q \Delta a_R^d}$
\\ \hline
 $\sqrt{|\mbox{Re}(\delta^d_{13})^2_{LR}|}$
& $3.3 \times 10^{-2} ~\tilde{m}_{\tilde{q}}$
&$\sim 2 \times 10^{-5}(\tilde A^d_a-\tilde A^d_b+\tilde A^d_{b'}
-\tilde A^d_c)\tilde{m}_{\tilde{q}}^{-1}$
\\ \hline
 $|(\delta^d_{23})_{LL,RR}|$
 & $8.2~ \tilde{m}_{\tilde{q}}^2$
 &$(LL)1.5 \times 10^{-2}\Delta a_L^q,(RR)4.7 \times 10^{-1}\Delta a_R^d$
 \\ \hline
 $|(\delta^d_{23})_{LR}|$
& $1.6 \times 10^{-2} ~\tilde{m}_{\tilde{q}}^2$
 &$\sim 5 \times 10^{-5}(\tilde A^d_a-\tilde A^d_b+\tilde A^d_{b'}
+0.1\tilde A^d_c)\tilde{m}_{\tilde{q}}^{-1}$
\\ \hline
\end{tabular}
\caption{\footnotesize{Experimental bounds on  $\delta$'s and their theoretical values 
in the $Q_6$ model,
 where 
$\tilde{m}_{\tilde{q}}$ denotes
$m_{\tilde{q}} /500$ GeV, and $\Delta a_{L,R}$
 and $\tilde A$ are given 
in (\ref{scalarmass}) and (\ref{Aterm}).}}
\label{deltas}
\end{center}
\end{table}

\subsection{The neutral Higgs bosons  and their mixing}
The scalar potential $V$ of the model consists
of the $\mu$ terms, the scalar soft masses, the $b$ terms
and the $D$ terms, and can be written as
\be
V &=&
m_{H_{+}^{u}}^{2}~(|\hat{H}_{+}^{0u}|^{2}+|\hat{H}_{-}^{0u}|^{2})+
m_{H_{+}^{d}}^{2}~(|\hat{H}_{+}^{0d}|^{2}+|\hat{H}_{-}^{0d}|^{2})
+m_{H_{3}^{u}}^{2}~|\hat{H}_{3}^{0u}|^{2}+
m_{H_{3}^{d}}^{2}~|\hat{H}_{3}^{0d}|^{2}\nn
& &+\frac{1}{8}(g_{Y}^{2}+
g_{2}^{2})(|\hat{H}_{+}^{0u}|^{2}+
|\hat{H}_{-}^{0u}|^{2} +|\hat{H}_{3}^{0u}|^{2}
-|\hat{H}_{+}^{0d}|^{2}-
|\hat{H}_{-}^{0d}|^{2} -|\hat{H}_{3}^{0d}|^{2} )^{2}\\
& +&\left[ b_{++} \hat{H}^{0u}_+ \hat{H}^{0d}_+
+b_{--}\hat{H}^{0u}_-\hat{H}^{0d}_-
+b_{+3}\hat{H}^{0u}_+\hat{H}^{0d}_3
+b_{3+}\hat{H}^{0u}_3\hat{H}^{0d}_+
+b_{33}\hat{H}^{0u}_3\hat{H}^{0d}_3 +h.c.\right],\no
\label{scalarp1}
\ee
where 
$g_{Y,2}$ are the gauge coupling constants
for the  $U(1)_{Y}$ and $SU(2)_{L}$ gauge groups,
and $H_\pm$'s are defined in (\ref{Hpm}).
As announced the scalar potential (\ref{scalarp1}) has the  $Z_2$
symmetry, where $H_{+}$'s and $H_{3}$'s are $Z_2$ even, and
$H_{-}$'s are $Z_2$ odd.
First we redefine the Higgs fields as 
$\tilde{H}^{0u,0d}_{+} = \hat{H}^{0u,0d}_{+} \exp- i \theta^{u,d}_{+},~
\tilde{H}^{0u,0d}_{3} = \hat{H}^{0u,0d}_{3} \exp -i \theta^{u,d}_{3},
$
and then define
\be
\phi^{u}_L &=& \cos\gamma^{u} \tilde{H}_3^{0u}
+\sin\gamma^{u} \tilde{H}_+^{0u},~
\phi^{u}_H = -\sin\gamma^{u} \tilde{H}_3^{0u}
+\cos\gamma^{u} \tilde{H}_+^{0u},
\label{phiLH}
\ee
where
\be
\cos\gamma^u &=& \frac{v_3^u}{\sqrt{(v_3^{u})^2+(v_+^{u})^2}},~
\sin\gamma^u = \frac{v_+^u}{\sqrt{(v_3^{u})^2+(v_+^{u})^2}},
\label{cosgamma}
\ee
and similarly for the down sector. As we see from (\ref{cosgamma}),
only $\phi^{u}_L$ and $\phi^{d}_L$ have a nonvanishing VEV, which we denote
by $
\sqrt{2}<\phi^{u,d}_L>
=\sqrt{(v^{u,d}_3)^2+(v^{u,d}_+)^2}=
v_{u,d}$.
The neutral light and heavy Higgs scalars of the MSSM are then given by
\be
\frac{1}{\sqrt{2}}( v + h-i X ) &=&( \phi^{d*}_L)\cos\beta
+ (\phi^{u}_L)\sin\beta,\\
\frac{1}{\sqrt{2}} (H+i A) &=&- (\phi^{d*}_L)\sin\beta
+ (\phi^{u}_L)\cos\beta,
\label{AHh}
\ee
where as in the MSSM
$v =\sqrt{v_u^2+v_d^2}$ and $\tan\beta=v_u/v_d$.

As in the case of the MSSM, the  couplings of $\phi^{u,d}_L$ are
flavor-diagonal, while the extra heavy fields
\be
\hat{H}^{0u,0d}_{-} &=& \phi_-^{u,d}=
(\varphi_-^{u,d}+i \chi_-^{u,d})/\sqrt{2}~,~
\phi^{u,d}_H=(\varphi_H^{u,d}+i \chi_H^{u,d})/\sqrt{2}
\label{phi-m}
\ee
can have flavor-changing couplings. The mass matrix  
for the $Z_2$-odd $\phi_-^{u,d}$ can be written as
\be
{\bf M}^2_- &=&\left(
\begin{array}{cccc}
m^2_{\phi_-^u} & 0 & b_-& -c_-\\
0 & m^2_{\phi_-^u}  & -c_-&-b_- \\
b_- & -c_- & m^2_{\phi_-^d} & 0\\
-c_- & -b_- & 0& m^2_{\phi_-^d}\\
\end{array}\right)~
\label{mass-m}
\ee
in the
$(\varphi_-^{u}, \chi_-^{u}, \varphi_-^{d}, \chi_-^{d})$  basis,
where $
m^2_{\phi_-^{u,d}} = m^2_{H_-^{u,d}}, b_-=\mbox{Re}(b_{--}), c_-=\mbox{Im} (b_{--})$.
The mass matrix  
for the $Z_2$-even fields is given by
\be
& &{\bf M}^2_H =\nn
& &\left(\begin{array}{ccccccc}
m_{\phi_H^u}^2&0  &  b_{HH}  & -c_{HH} & m_{\phi_{HL}^u}^2/c_\beta
 &0 &0\\
0  &m_{\phi_H^u}^2  &  -c_{HH}  & -b_{HH} &
0  & m_{\phi_{HL}^u}^2/c_\beta   &0\\
b_{HH}  &-c_{HH}  &  m_{\phi_H^d}^2  & 0 &
 -m_{\phi_{HL}^d}^2/s_\beta   & 0 &0\\
-c_{HH}  &-b_{HH}  &  0  &  m_{\phi_H^d}^2  &
0  &  m_{\phi_{HL}^d}^2/s_\beta & 0\\
 m_{\phi_{HL}^u}^2/c_\beta  &0  &   -m_{\phi_{HL}^d}^2/s_\beta  & 
0 & m_{\phi_L^u}^2+m_{\phi_L^d}^2  +s_{2\beta}^2 M_Z^2
 & 0 &-c_{2\beta}s_{2\beta} M_Z^2\\
0  & m_{\phi_{HL}^u}^2/c_\beta    &  0  &  
m_{\phi_{HL}^d}^2/s_\beta   &0  & 
m_{\phi_L^u}^2+m_{\phi_L^d}^2 &0\\
0  & 0  &  0  &  
0 &-c_{2\beta}s_{2\beta} M_Z^2  & 0 &c_{2\beta}^2 M_Z^2\\
\end{array}\right)\nn
\label{mass-H-A-h}
\ee
in the $(\varphi_H^u, \chi_H^u,   \varphi_H^d, \chi_H^d,  H, A, h)$ basis, where
$m_{\phi_H^u}^2 =\hat{m}_{\phi_H^u}^2-c_\beta M_Z^2/2~,~
m_{\phi_H^d}^2 =\hat{m}_{\phi_H^d}^2+c_\beta M_Z^2/2~,~
c_{a\beta}=\cos a\beta~,~s_{a\beta}=\sin a\beta$,
\be
\hat{m}^2_{\phi_H^{u,d}} &=&
m^2_{H_+^{u,d}}\cos^2\gamma^{u,d}+
m^2_{H_3^{u,d}}\sin^2\gamma^{u,d}~,~
m^2_{\phi_L^{u,d}} =
m^2_{H_+^{u,d}}\sin^2\gamma^{u,d}+
m^2_{H_3^{u,d}}\cos^2\gamma^{u,d}~,\nn
m^2_{\phi_{HL}^{u,d}} &=&
\frac{1}{2}\sin 2\gamma^{u,d}
(m^2_{H_+^{u,d}}-
m^2_{H_3^{u,d}}),\nn
b_{HH}+i c_{HH} &=&b_{++} e^{-i(\theta_+^u+\theta_+^d)} 
\cos\gamma^u \cos\gamma^d-
b_{+3} \cos\gamma^u \sin\gamma^d e^{-i(\theta_+^u+\theta_3^d)}\nn 
& -&
b_{3+} \sin\gamma^u \cos\gamma^d e^{-i(\theta_3^u+\theta_+^d)} +
b_{33} \sin\gamma^u \sin\gamma^d e^{-i(\theta_3^u+\theta_3^d)} ~.
\label{mandb}
\ee
All the parameters in the mass matrices 
(\ref{mass-m}) and (\ref{mass-H-A-h}) are  real, and
the mass parameters and $\gamma^{u,d}$ are defined in (\ref{cosgamma}).
In \cite{Kifune:2007fj} it was assumed that $m_{\phi_{HL}^{u,d}}^2$
(which express the mixing among the MSSM and extra heavy Higgs fields) are 
small compared with other mass parameters such as 
$m_{\phi_{H}^{u,d}}^2$. Under this assumption the mass matrix squared (\ref{mass-H-A-h})
goes over to the one given in \cite{Kifune:2007fj}.

\section{$B^0-\bar{B}^0$ mixing via heavy neutral Higgs bosons}
As a last task  we investigate 
signatures of new physics contributions to the non-diagonal
matrix element of the effective hamiltonians of the neutral
meson systems $M_{12}$.
We will see that not only the contributions from the supersymmetry breaking
sector (as we have found in section 2.2), but also those from the flavor-changing
neutral Higgs exchanges are real,  and that despite being real 
the new contributions can create a new mixing phase.

The total matrix element  $M_{12}$ can be written as
\be
M_{12} &= &M_{12}^{SM} + M_{12}^{\rm new} =  M_{12}^{SM} \cdot \Delta \, ,
\label{M12}
\ee
and we follow \cite{LN2006} to parameterize new physics effects
in the observables
$\Delta M_s$, $\Delta \Gamma_s$ and the flavor specific CP
asymmetry $a_{fs}^s$ in terms of the complex number 
$\Delta_s = |\Delta_s| e^{i \phi^{\Delta}_s}$:
\be
\dm_s  & = & 2|M_{12}^{SM} \cdot \Delta_s | =\dm_s^{\rm SM} \,  |\Delta_s|
~,
\Delta \Gamma_s  = 2 |\Gamma_{12}^s|
     \, \cos \left( \phi_s^{\rm SM} + \phi^\Delta_s \right),
\label{bounddg}
\\
\frac{\Delta \Gamma_s}{\Delta M_s}
&= &
 \frac{|\Gamma_{12}^s|}{|M_{12}^{\rm SM,s}|}
\cdot \frac{\cos \left( \phi_s^{\rm SM} + \phi^\Delta_s \right)}{|\Delta_s|},
a_{\rm fs}^s
=
 \frac{|\Gamma_{12}^s|}{|M_{12}^{\rm SM,s}|}
\cdot \frac{\sin \left( \phi_s^{\rm SM} + \phi^\Delta_s \right)}{|\Delta_s|}.
\label{boundafs} 
\ee
The SM values are given e.g.  in \cite{LN2006}, in which the results of
\cite{BBD06,BBGLN98,BBGLN02,BBLN03,rome03} are used.
For the present 
model with the CKM parameters given in (\ref{ckm-parameters}) we have 
\cite{Kifune:2007fj}
\footnote{The phase for $M^{\rm SM,s}_{12}$
given in \cite{Kifune:2007fj} $-i 0.0035$
should be replaced by $ -i 0.035$.}:
\be
 2~M^{\rm SM,s}_{12}&=&
20.1(1\pm 0.40)\exp (-i 0.035)
~\mbox{ps}^{-1},
\label{m12s}\\
2~M^{\rm SM,d}_{12}&=&
0.56(1\pm 0.45)\exp (i 0.77)
~\mbox{ps}^{-1},\\
\label{m12d}
\phi_s^{\rm SM}& = &(4.2\pm 1.4) \cdot 10^{-3}~\mbox{rad}~,~
 \Delta \Gamma_s^{\rm SM} =
 0.096\pm0.039 ~\mbox{ps}^{-1},
 \label{phis}\\
 a_{fs}^{\rm SM,s} &=& (2.06\pm 0.57) \cdot 10^{-5},
\label{afs}
\ee
where the errors are dominated by the uncertainty in the 
decay constants $f_{B}$.
The corresponding experimental values are given by \cite{HFAG}
\be
\Delta M_s^{\rm exp} &=&17.78\pm 0.12
~\mbox{ps}^{-1},\label{m12sexp}\\
\Delta M_d^{\rm exp} &=&0.507\pm 0.005~\mbox{ps}^{-1},\label{m12dexp}\\
a_{fs}^{\rm exp,s} (=a_{sl}^{\rm exp,s})
&=&-0.0037\pm 0.0094 \, ,
\label{afsexp}
\ee
and for $\Delta \Gamma_s$ and  $\phi_s=\phi_s^{\rm SM} + \phi^\Delta_s$
there are two regions \cite{HFAG}:
\be
\phi_s^{\rm exp} &=& -2.36
\mbox{\small $\begin{array}{c}
+0.37 \\ -0.29
\end{array}$}~\mbox{rad}~,~
\Delta \Gamma_s^{\rm exp}=-0.154
\mbox{\small $\begin{array}{c}
+0.070 \\ -0.054
\end{array}$}~\mbox{ps}^{-1}~,
\label{phi1}\\
\phi_s^{\rm exp} &=&-0.77
\mbox{\small $\begin{array}{c}
+0.29 \\ -0.37
\end{array}$}~\mbox{rad}~,~
\Delta \Gamma_s^{\rm exp}=0.154
\mbox{\small $\begin{array}{c}
+0.054 \\ -0.070
\end{array}$}~\mbox{ps}^{-1}~.
\label{phi2}
\ee
The above experimental values are $2.2 \sigma$  away from the SM
prediction (\ref{phis}) \cite{LN2006},
which may indicate a possible
existence of new physics \cite{LN2006,HFAG,CKMfitter,UTfit}. 
With this in mind, we proceed with our investigation
on possible new effects.

The Lagrangian  that describes the mixing of $B^0$ and $\bar{B}^0$
(also that of $K^0$ and $\bar{K}^0$)
is given by
\be
{\cal L}_{FCNC}
&=&-\left[~Y_{ij}^{dH}\phi_H^{d}+Y_{ij}^{d-}
\phi^{d}_-~\right]^*
\overline{d}_{iL}' d_{jR}'+h.c.~,
\label{Lfcnc}
\ee
where $d'$'s are  mass eigenstates,
the Higgs fields are defined in (\ref{phiLH}) and (\ref{phi-m}), and \cite{Kifune:2007fj}
\be
{\bf Y}^{dH}
&\simeq&\frac{1}{\tan\gamma^d\cos\beta}\left( \begin{array}{ccc}
6.63\times 10^{-5} &  
8.26\times 10^{-5} &2.80\times 10^{-4}
\\-6.224\times 10^{-5}
 &3.74\times 10^{-4}
&3.37\times 10^{-4}
  \\4.10\times 10^{-3}
 &-6.01\times 10^{-3}  &2.52\times 10^{-3}
\end{array}\right)\nn
& &-\frac{\tan\gamma^d}{\cos\beta}\left( \begin{array}{ccc}
1.37\times 10^{-5} &  
1.13\times 10^{-4} &7.56\times 10^{-5}
\\1.98\times 10^{-5}
 &-1.88\times 10^{-4}
&-3.72\times 10^{-4}
  \\1.67\times 10^{-3}
 &6.61\times 10^{-3}  &0.0131
\end{array}\right)~,
\label{YdH}
\ee
\be
{\bf Y}^{d-}
&\simeq&\frac{\exp i(2\theta^d_3-\theta^d_+)}
{\sin\gamma^d\cos\beta}\left( \begin{array}{ccc}
0 &  
-2.53\times 10^{-4}
& -4.72\times 10^{-4}
\\-2.22\times 10^{-4}
 & 0 &-1.04\times 10^{-4}
  \\7.46\times 10^{-3}
   &-1.89\times 10^{-3}
   &  0   \end{array}\right).
   \label{Ydm}
   \ee
 The phases appearing in the Yukawa matrices above are given in (\ref{vev1}).
Given the FCNC interactions (\ref{Lfcnc}) we are now able to compute
the extra contribution $M_{12}^{\rm new}$.
To this end we  need to compute the inverse 
of the mass matrices squared (\ref{mass-m}) and (\ref{mass-H-A-h}),
which we denote by $\Delta^-$ and $\Delta^H$, respectively.
The elements of $\Delta$'s relevant to our purpose are:
\be
\Delta^- _{\varphi_-^d -\varphi_-^d} &=& \Delta^- _{\chi_-^d -\chi_-^d}
=\frac{m^2_{\phi_-^u}}{(\bar {M}_-^2)^2}\equiv \frac{1}{(M_-^d)^2}
~,~\Delta^- _{\varphi_-^d -\chi_-^d}=0,
\label{Delta-m}\\
\Delta_{\varphi_H^d-\varphi_H^d} &=& \Delta_{\chi_H^d-\chi_H^d}=\frac{ (m^2_{\phi_{HL}^{u}})^2/\cos^2\beta-
 m^2_{\phi_{H}^{u}}m_H^2}{(\bar{M}_H^2)^3}
 \equiv \frac{1}{(M_H^d)^2},
 \label{Delta-phi-phi}\\
 \Delta_{\varphi_H^d-\chi_H^d} &=&0,
  \label{Delta-phi-chi}
 \ee
 where $( \bar{M_-}^2)^4=\det {\bf M}^2_-$ and 
$( \bar{M}_H^2)^6 =\det {\bf M}^2_H/\cos^2 2\beta M_Z^2$ and 
$m_H^2=m^2_{\phi^u_L}+m^2_{\phi^d_L}$.
The mass parameters appearing in 
(\ref{Delta-m}) -(\ref{Delta-phi-chi}) are defined in (\ref{mass-m}) and (\ref{mass-H-A-h}).
The fact that $\Delta_{\varphi^d -\varphi^d}=\Delta_{\chi^d -\chi^d}$
and $\Delta _{\varphi^d -\chi^d}=0$ has an important consequence that
although CP is explicitly broken by the 
b terms in  the supersymmetry breaking sector,
the new contribution to $M_{12}$ is real, as in the case of the
contribution from the supersymmetry breaking sector.
Therefore, the new contributions $M^{\rm new}_{12}$ 
from the $\varphi$ and $\chi$ exchanges take the form \cite{Kifune:2007fj}
\be
M^{\rm new,K}_{12}
  &=&2
<\overline{K}^0| C_{K}
\overline{s}_R^\alpha d_L^\alpha 
\overline{s}_L^\beta d_R^\beta
|K^0> 
\simeq  0.56~C_{K}\mbox{GeV}^3,\nn
C_{K}
&=&   \left[  
(Y_{sd}^{dH})^*Y_{ds}^{dH}/(\cos\beta M_H^d)^2+
(Y_{sd}^{d-})^* Y_{ds}^{d-})/( \cos\beta M_-^d)^2 \right],
\label{K12}
\\
M^{\rm new, d}_{12}
  &=&2
<\overline{B}^0_d|
C_{d}(m_b)\overline{b}_R^\alpha d_L^\alpha 
\overline{b}_L^\beta d_R^\beta
|B^0_d>
\simeq  0.36~C_{d}(m_b)\mbox{GeV}^3,\nn
C_{d}(m_b)
&=&   \eta(m_b) \left[  
(Y_{bd}^{dH})^*Y_{db}^{dH}/(\cos\beta M_H^d)^2+
(Y_{bd}^{d-})^* Y_{db}^{d-})/(\cos\beta M_-^d)^2 \right],
\label{d12}\\
M^{\rm new,s}_{12}
  &=&2
<\overline{B}^0_s|
C_{s}(m_b)\overline{b}_R^\alpha s_L^\alpha 
\overline{b}_L^\beta s_R^\beta 
|B^0_s>
\simeq  0.58~C_{s}(m_b)\mbox{GeV}^3,\nn
C_{s}(m_b)
&=&   \eta(m_b) \left[  
(Y_{bs}^{dH})^* Y_{sb}^{dH}/(\cos\beta M_H^d)^2+
(Y_{bs}^{d-})^* Y_{sb}^{d-}/(\cos\beta M_-^d)^2 \right],
\label{s12}
\ee
where  
$\eta(m_b)\simeq 2.0 $ is the one-loop QCD correction, $Y$'s are elements of the
Yukawa matrices (\ref{YdH}) and (\ref{Ydm}). 
The matrix elements (\ref{K12})-(\ref{s12}) basically suffer
from the same size of the uncertainties as (\ref{m12s}) and (\ref{m12d}).
In the following calculations we impose  
the constraints
\be
0.6 &< &\Delta M_{d,s}/\Delta M_{d,s}^{\rm exp } < 1.4~,~
2 | (M_K^{\rm new})_{12} | < \Delta M_{K}^{\rm exp } \simeq ~3.49\times 10^{-15}~\mbox{GeV},\nn
|\phi_d^\Delta | & < &  0.17~\mbox{rad}.
\label{const2}
\ee
($\phi_d^\Delta$ is an analog 
of $\phi_s^\Delta$   for $B_d$.)
 \begin{figure}[htb]
\includegraphics*[width=0.6\textwidth]{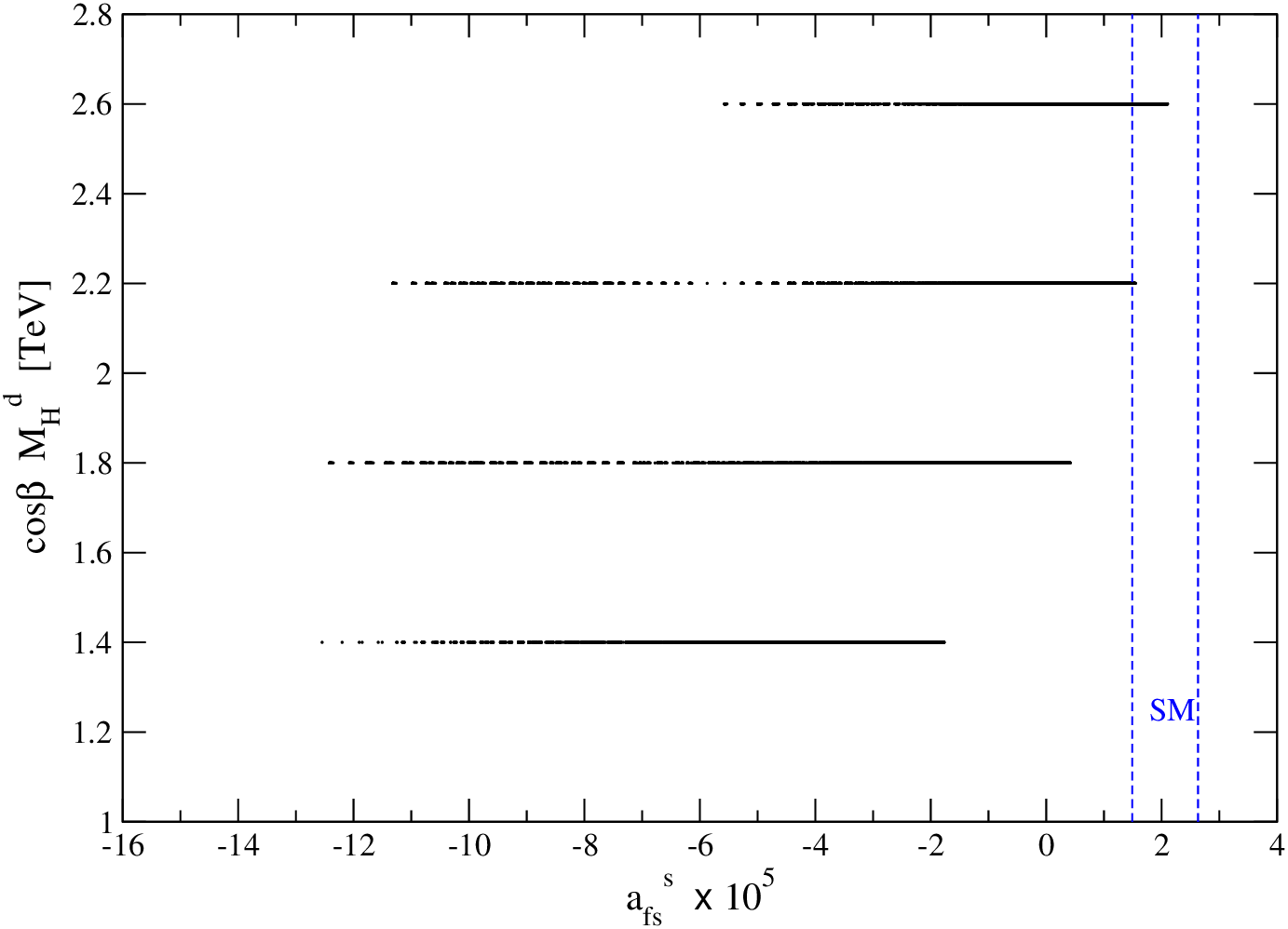}
\caption{\label{fcnc}\footnotesize
 The prediction of  the CP asymmetry $a_{fs}^s \times 10^5$ for
 different values of the Higgs mass $\cos\beta  M_{H}^d$ [TeV].
The SM value is between two blue lines. }
\end{figure}

\vspace{0.2cm}\noindent
(i) \underline{$\phi_s^\Delta $}\\
We first compute $\phi_s^\Delta $. 
To this end we include all the contributions;
the contributions from the  $\varphi $ and $\chi$ exchanges and 
those from the soft supersymmetry breaking terms, where we assume that
the later contributions can be freely chosen by varying the 
$a_{L,R}$'s and $A_{ij}$'s defined in (\ref{scalarmass}) and (\ref{Aterm}).
We find:
\be
-0.018 \lsim \phi_s^{\rm SM} + \phi^\Delta_s
\lsim 0.012 ~~\mbox{and}~~
-0.023 \lsim  \phi^\Delta_s \lsim 0.009.
\label{phis3}
\ee
(If only the Higgs exchanges are taken into account,
we find -$0.015 \lsim \phi_s^{\rm SM} + \phi^\Delta_s
\lsim 0.007$.)
So, if the evidence for a new phase (\ref{phi1})
or (\ref{phi2}) were confirmed, not only the SM, but also the present supersymmetric model
might run into a serious problem \footnote{A similar conclusion 
has been reached in \cite{Gorbahn:2009pp}
for the MSSM with large $\tan\beta$ and the Minimal Flavor Violation assumption.}.\\

\vspace{0.2cm}\noindent
(ii) \underline{$a_{fs}^s$}\\
Using (\ref{boundafs}) we next compute
$a_{fs}^s/a_{fs}^{\rm SM,s}=
\sin (\phi_s^{\rm SM}+\phi_s^{\Delta})/\sin \phi_s^{\rm SM}|\Delta_s|$.
First we consider only the contributions from 
the Higgs  exchanges, where for a given $\cos\beta M_H^d$ 
we vary the Higgs mixing angle  $\gamma^d$ (\ref{cosgamma}) and $r=M_-^d/M_H^d$  so as to satisfy
the constraints (\ref{const2}).
The result is plotted in Fig.~\ref{fcnc}, where we varied
$\cos\beta M_H^d$ from $1.2$ (the smallest allowed value) to $2.6$ TeV.
The SM value (\ref{afs}) is between to blue vertical lines.
If all thre contributions are included, we find
\be
-13 &\lsim & a_{fs}^s\times 10^5 \lsim 7.
\ee
(The experimental value is given in  (\ref{afsexp}).)

\begin{figure}[htb]
\includegraphics*[width=0.45\textwidth]{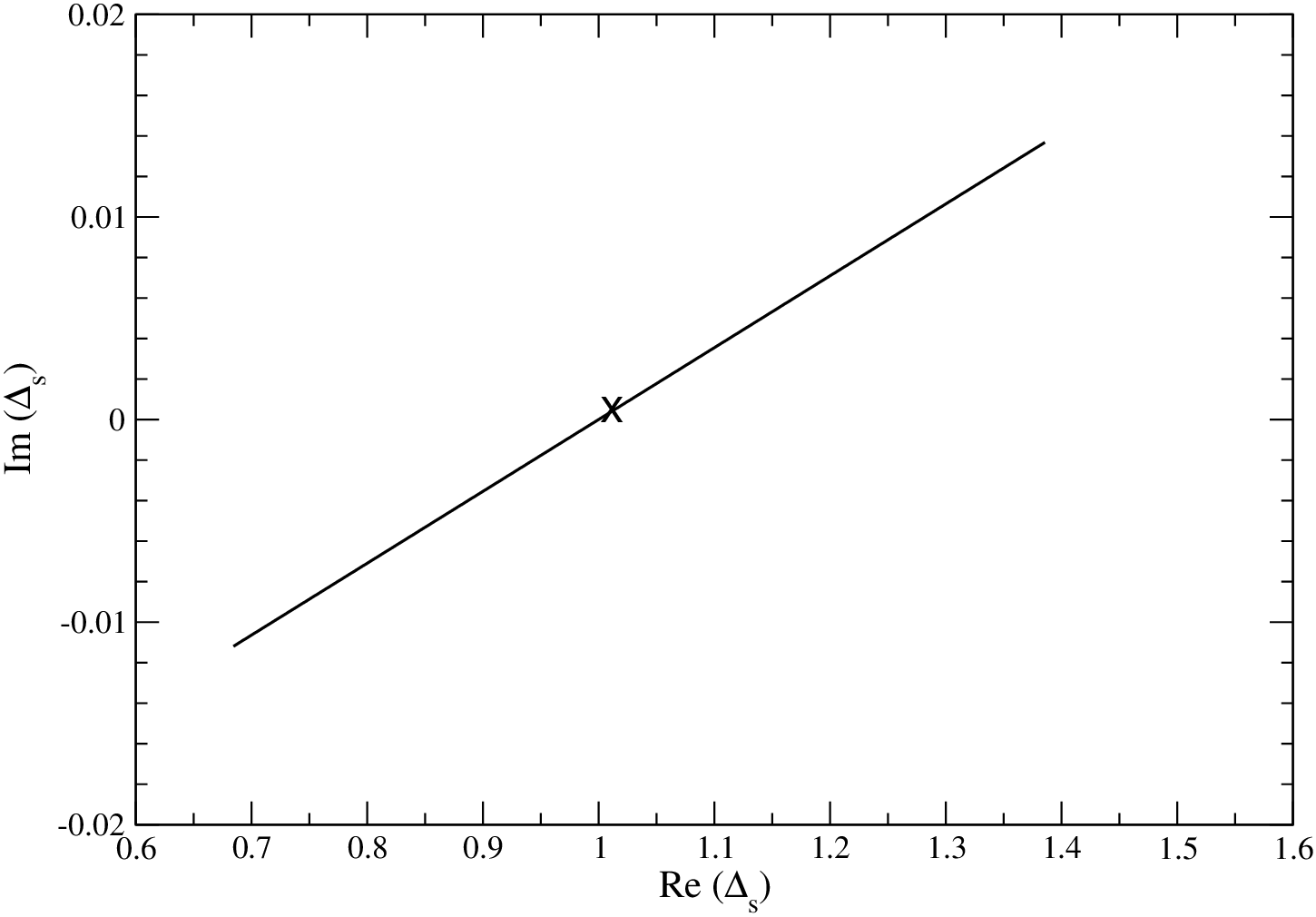}
\includegraphics*[width=0.45\textwidth]{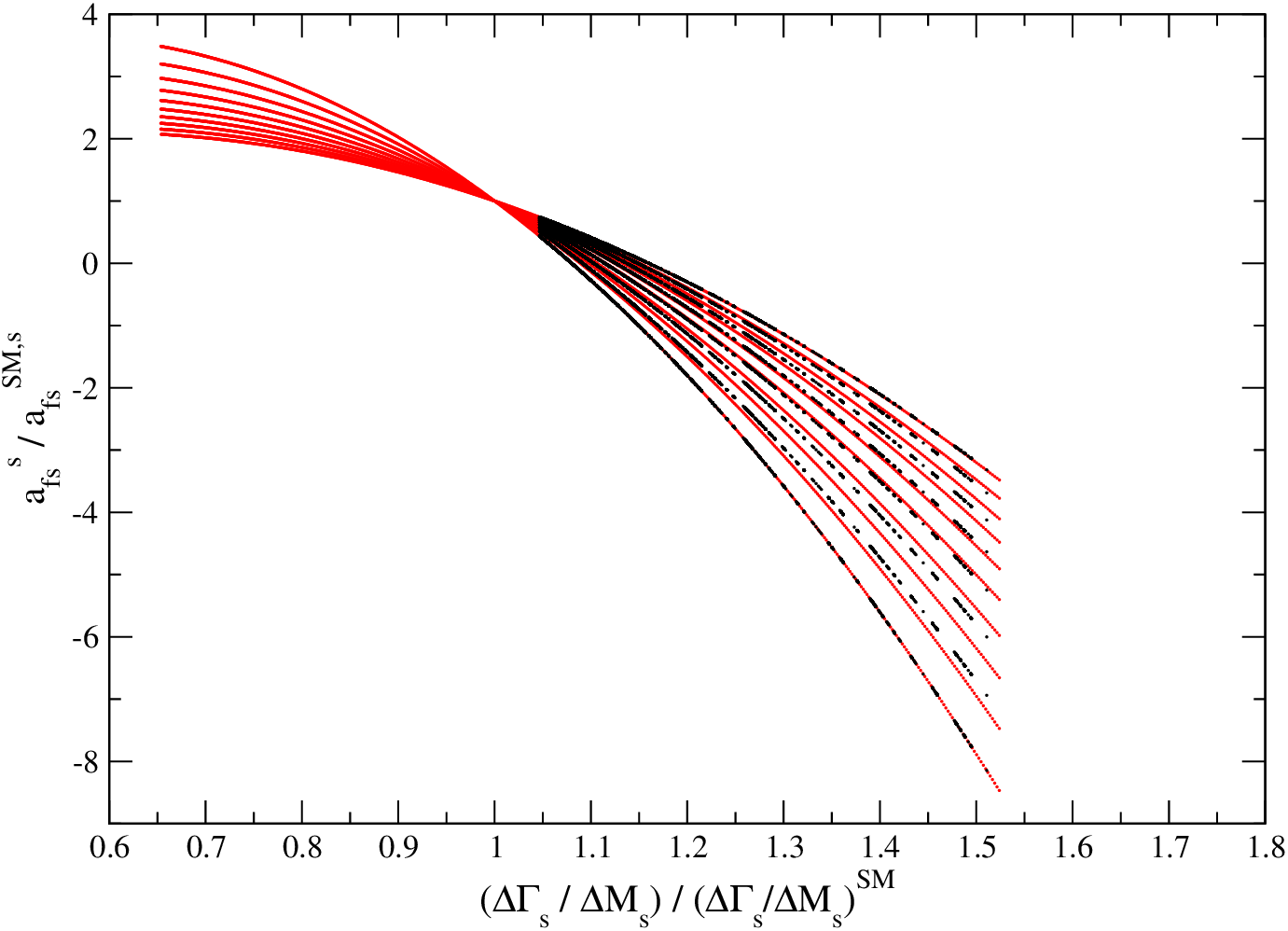}
\caption{\label{re-im}\footnotesize
Right:  The prediction in the $(\Delta \Gamma_s/\Delta M_s)/
(\Delta \Gamma_s/\Delta M_s)^{SM}-
(a_{fs}^s)/(a_{fs}^s)^{SM} $ plane.
The black points are those 
without the contribution from
the soft-supersymmetry breaking terms.
The red points are obtained
by including the contributions 
coming from both the  Higgs exchanges
 and  soft-supersymmetry breaking terms.
Left: The prediction in the $Re (\Delta_s)-Im (\Delta_s)$ plane,
where  the contributions 
coming from both the  Higgs  exchanges  
 and  soft-supersymmetry breaking terms are included.
The cross denotes the SM point.
The Higgs mass $\cos\beta  M_{H}^d$ is varied  from $1.2$ to $3.0$ GeV
for both panels.
 }
\end{figure}

\vspace{0.2cm}\noindent
(iii) \underline{$(\Delta \Gamma_s/\Delta M_s)-a_{fs}^s$}\\
 The prediction of $(a_{fs}^s)/(a_{fs})^{\rm SM}$
 against $(\Delta \Gamma_s/\Delta M_s)/(\Delta \Gamma_s/\Delta M_s)^{\rm SM}$
 is plotted in Fig.~ \ref{re-im} (right).
The contribution only from 
the  Higgs exchanges is indicated by black.
In this area $a_{fs}^s$ is mostly negative and its size may become
one order of magnitude larger than the SM value.

\vspace{0.2cm}\noindent
(iv) \underline{$\Delta_s$}\\
 The prediction in the $Re (\Delta_s)-Im (\Delta_s)$
 is plotted in Fig.~ \ref{re-im} (left),
 where the cross denotes the SM point.
All the contribution  are taken into account.

\section{Conclusion}
We  considered a supersymmetric extension of the SM
based on the discrete $Q_6$ family symmetry, which has been recently proposed in 
\cite{Babu:2004tn,Kajiyama:2005rk}, and  investigated the extra contribution to
$M_{12}$, which we denoted by $M_{12}^{\rm new}$.
We assumed that CP is explicitly, but softly broken only
by the b terms in the soft supersymmetry breaking sector.
Therefore, all  other parameters of the model are real,
which  is consistent with renormalizability \cite{Kifune:2007fj}.
There are two origins for the contribution to $M_{12}^{\rm new}$;
from the supersymmetry breaking sector
and from the exchange of the flavor-changing neutral Higgs bosons.
We found that both contributions are real, and 
that nevertheless we obtain an observable  difference in the CP 
violation.
We focus our attention on  the extra $B_s$-mixing phase $\phi_s^\Delta$ 
and the flavor-specific CP asymmetry $a_{fs}^s$,
because they are accidentally small $\sim O(10^{-3})$ and  
$\sim O(10^{-5})$, respectively, in the SM.
We found that $a_{fs}^s$ in our model is mostly
negative and can be indeed one order of magnitude larger the SM value
in size. Our results Fig.~\ref{fcnc} and \ref{re-im}, which  are
consistent with the  current experimental value (\ref{phi2}), 
can be experimentally tested  e.g. by LHCb near future \cite{lhcb}.

\vspace*{5mm}
J.~K. is partially supported by a Grant-in-Aid for Scientific
Research (C) from Japan Society for Promotion of Science (No.18540257).
A.L. would like to thank  M.
      Sedlmeier (AAA, University of Regensburg), DFG  and the theory group of
     Kanazawa  for
    the  financial support for the stay in Kanazawa in 2008 during
      that part of this work was performed.


\begin{thebibliography}{99}


\bibitem{Paulini:2009ri}
  M.~Paulini,
  arXiv:0906.3786 [hep-ex].

\bibitem{LN2006}
 A.~Lenz and U.~Nierste,
  JHEP {\bf 0706} (2007) 072
  [arXiv:hep-ph/0612167].
  
  \bibitem{DMCDF}
A.~Abulencia {\it et al.}  [CDF Collaboration],
  Phys.\ Rev.\ Lett.\  {\bf 97} (2006) 242003
  [arXiv:hep-ex/0609040].
  
  
\bibitem{DMD0}
  V.~M.~Abazov {\it et al.}  [D0 Collaboration],
  Phys.\ Rev.\ Lett.\  {\bf 97} (2006) 021802
  [arXiv:hep-ex/0603029].
  
\bibitem{HFAG}
E.~Barberio {\it et al.},
  arXiv:0808.1297 [hep-ex].


\bibitem{FDN}
  I.~Dunietz, R.~Fleischer and U.~Nierste,
  Phys.\ Rev.\  D {\bf 63} (2001) 114015
  [arXiv:hep-ph/0012219].

\bibitem{ELBA}
  A.~Lenz,
  Nucl.\ Phys.\ Proc.\ Suppl.\  {\bf 177-178} (2008) 81
  [arXiv:0705.3802 [hep-ph]].
  
  \bibitem{Aaltonen:2007he}
 T.~Aaltonen {\it et al.}  [CDF collaboration],
  Phys.\ Rev.\ Lett.\  {\bf 100} (2008) 121803
  [arXiv:0712.2348 [hep-ex]];
%
  T.~Aaltonen {\it et al.}  [CDF Collaboration],
  Phys.\ Rev.\ Lett.\  {\bf 100} (2008) 161802
  [arXiv:0712.2397 [hep-ex]];
%
  CDF note 9458.
  
  \bibitem{2008fj}
  V.~M.~Abazov {\it et al.}  [D0 Collaboration],
  Phys.\ Rev.\ Lett.\  {\bf 101} (2008) 241801
  [arXiv:0802.2255 [hep-ex]];
%
  D0 note 5933.

  \bibitem{CKMfitter}
 A.~Hocker, H.~Lacker, S.~Laplace and F.~Le Diberder,
  Eur.\ Phys.\ J.\  C {\bf 21} (2001) 225
  [arXiv:hep-ph/0104062];
%
  J.~Charles {\it et al.}  [CKMfitter Group],
  Eur.\ Phys.\ J.\  C {\bf 41} (2005) 1
  [arXiv:hep-ph/0406184];
%
http://ckmfitter.in2p3.fr/

 \bibitem{UTfit}
   M.~Bona {\it et al.},
  arXiv:0906.0953 [hep-ph],
%
  http://www.utfit.org/


\bibitem{asldirect}
  V.~M.~Abazov {\it et al.}  [D0 Collaboration],
  arXiv:0904.3907 [hep-ex].

\bibitem{asldimuon}
  V.~M.~Abazov {\it et al.}  [D0 Collaboration],
  Phys.\ Rev.\  D {\bf 74} (2006) 092001
  [arXiv:hep-ex/0609014];
%
CDF note 9015.


\bibitem{BBD06}
 M.~Beneke, G.~Buchalla and I.~Dunietz,
  Phys.\ Rev.\  D {\bf 54} (1996) 4419
  [arXiv:hep-ph/9605259].



\bibitem{BBGLN98} 
M.~Beneke, G.~Buchalla, C.~Greub, A.~Lenz and U.~Nierste,
  Phys.\ Lett.\  B {\bf 459} (1999) 631
  [arXiv:hep-ph/9808385].

\bibitem{BBGLN02} 
 M.~Beneke, G.~Buchalla, C.~Greub, A.~Lenz and U.~Nierste,
  Nucl.\ Phys.\  B {\bf 639} (2002) 389
  [arXiv:hep-ph/0202106].

\bibitem{BBLN03}
M.~Beneke, G.~Buchalla, A.~Lenz and U.~Nierste,
  Phys.\ Lett.\  B {\bf 576} (2003) 173
  [arXiv:hep-ph/0307344].



\bibitem{rome03}
M.~Ciuchini, E.~Franco, V.~Lubicz, F.~Mescia and C.~Tarantino,
  JHEP {\bf 0308} (2003) 031
  [arXiv:hep-ph/0308029].
  
  
     \bibitem{Babu:2004tn}
  K.~S.~Babu and J.~Kubo,
  Phys.\ Rev.\  D {\bf 71} (2005) 056006
  [arXiv:hep-ph/0411226].

\bibitem{Kajiyama:2005rk}
 E.~Itou, Y.~Kajiyama  and J.~Kubo,
  Nucl.\ Phys.\  B {\bf 743} (2006) 74
  [arXiv:hep-ph/0511268].

\bibitem{Kifune:2007fj}
  N.~Kifune, J.~Kubo and A.~Lenz,
  Phys.\ Rev.\  D {\bf 77} (2008) 076010
  [arXiv:0712.0503 [hep-ph]].
                                 
 \bibitem{Kubo:2003iw}
  J.~Kubo, A.~Mondragon, M.~Mondragon and E.~Rodriguez-Jauregui,
  Prog.\ Theor.\ Phys.\  {\bf 109} (2003) 795
  [Erratum-ibid.\  {\bf 114} (2005) 287]
  [arXiv:hep-ph/0302196];
  J.~Kubo,
  Phys.\ Lett.\  B {\bf 578} (2004) 156
  [Erratum-ibid.\  B {\bf 619} (2005) 387]
  [arXiv:hep-ph/0309167].
  
  \bibitem{Mondragon:2008gm}
  A.~Mondragon, M.~Mondragon and E.~Peinado,
  arXiv:0805.3507 [hep-ph].
                                
\bibitem{Araki:2008rn}
  T.~Araki and J.~Kubo,
  arXiv:0809.5136 [hep-ph].

      \bibitem{Hall:1985dx}
    L.~J.~Hall, V.~A.~Kostelecky and S.~Raby,
  Nucl.\ Phys.\  B {\bf 267} (1986) 415.
  
\bibitem{Gabbiani:1988rb}
  F.~Gabbiani and A.~Masiero,
  Nucl.\ Phys.\  B {\bf 322} (1989) 235.      
  

\bibitem{Gorbahn:2009pp}
  M.~Gorbahn, S.~Jager, U.~Nierste and S.~Trine,
  arXiv:0901.2065 [hep-ph].
  
\bibitem{lhcb}
  C.~Langenbruch, U.~Uwer and S.~Hansmann-Menzemer,
CERN-LHCb-2009-028;
%
 N.~Brook, N.~Cottingham, R.~Lambert, F.~ Muheim, J.~Rademacker, P.~Szczypka and Y.~Xie,
CERN-LHCb-2007-054;
U.~Uwer, talk given at Moriond 2009, 
http://indico.in2p3.fr/getFile.py/access?contribId=13\&sessionId=2\&resId=0\&materialId=slides\&confId=1399

\end{thebibliography}
\end{document}